\documentclass[a4paper,11pt]{article}
\usepackage{pos}

\title{Comparing meson-meson and diquark-antidiquark creation operators for a $\bar b \bar b u d$ tetraquark}
\ShortTitle{Comparing $BB$ and $DD$ creation operators for a $\bar b \bar b u d$ tetraquark}

\author*[a,b]{Marc Wagner}
\author[c]{Pedro Bicudo}
\author[d]{Antje Peters}
\author[a]{Sebastian Velten}

\affiliation[a]{Johann Wolfgang Goethe-Universit\"at Frankfurt am Main,\\
  Institut f\"ur Theoretische Physik, Max-von-Laue-Stra{\ss}e 1, D-60438 Frankfurt am Main, Germany}

\affiliation[b]{Helmholtz Research Academy Hesse for FAIR,\\
  Campus Riedberg, Max-von-Laue-Stra{\ss}e 12, D-60438 Frankfurt am Main, Germany}

\affiliation[c]{CeFEMA, Dep.\ F\'{\i}sica, Instituto Superior T\'ecnico,\\ Universidade de Lisboa, Av.\ Rovisco Pais, 1049-001 Lisboa, Portugal}

\affiliation[d]{Westf\"alische Wilhelms-Universit\"at M\"unster,\\
  Institut f\"ur Medizinische Psychologie und Systemneurowissenschaften, Von-Esmarch-Straße 52, D-48149 M\"unster, Germany}

\emailAdd{mwagner@itp.uni-frankfurt.de}
\emailAdd{bicudo@tecnico.ulisboa.pt}
\emailAdd{antje.peters@uni-muenster.de}
\emailAdd{velten@itp.uni-frankfurt.de}

\abstract{We compare two frequently discussed competing structures for a stable $\bar b \bar b u d$ tetraquark with quantum numbers $I(J^P) = 0(1^+)$ by considering a meson-meson as well as a diquark-antidiquark creation operator. We treat the heavy antiquarks as static with fixed positions and find diquark-antidiquark dominance for $\bar b \bar b$ separations $r \ltapprox 0.2 \, \text{fm}$, while for $r \gtapprox 0.5 \, \text{fm}$ the system essentially corresponds to a pair of $B$ mesons. For the meson-meson to diquark-antidiquark ratio of the tetraquark we obtain around $58\%/42\%$.}

\FullConference{%
 The 38th International Symposium on Lattice Field Theory, LATTICE2021
  26th-30th July, 2021
  Zoom/Gather@Massachusetts Institute of Technology
}

\newcommand{\gtapprox}{\raisebox{-0.3ex}{$\,\stackrel{>}{\scriptstyle\sim}\,$}}
\newcommand{\ltapprox}{\raisebox{-0.3ex}{$\,\stackrel{<}{\scriptstyle\sim}\,$}}

\begin{document}

\maketitle

% ********************

\section{Introduction}

Anti-heavy-anti-heavy-light-light tetraquarks $\bar{Q} \bar{Q} q q$ are expected to be hadronically stable, if the antiquarks are sufficiently heavy (see e.g.\ Refs.\ \cite{Ader:1981db,Ballot:1983iv,Lipkin:1986dw,Heller:1986bt,Carlson:1987hh}). This was confirmed numerically by lattice QCD computations using the Born-Oppenheimer approximation \cite{Bicudo:2012qt,Brown:2012tm,Bicudo:2015vta,Bicudo:2015kna,Bicudo:2016ooe} as well as by full lattice QCD computations using four quarks of finite mass \cite{Francis:2016hui,Francis:2018jyb,Junnarkar:2018twb,Leskovec:2019ioa,Hudspith:2020tdf}. 

In this work (see also Ref.\ \cite{Bicudo:2021qxj}) we continue our Born-Oppenheimer based lattice QCD studies and explore the structure of a theoretically predicted $\bar b \bar b u d$ tetraquark with quantum numbers $I(J^P) = 0(1^+)$. In particular we try to clarify, whether it resembles a meson-meson system or rather a diquark-antidiquark system. Experimentally, this tetraquark has not yet been observed, but its discovery potential is discussed in Refs.\ \cite{Ali:2018ifm,Ali:2018xfq}.

% ********************

\section{\label{SEC001}Basic principle of our approach and summary of previous work}

The Born-Oppenheimer approximation \cite{Born:1927,Braaten:2014qka} can be used to study $\bar b \bar b q q $ tetraquarks in a two step approach. In a first step, one treats the heavy $\bar b$ quarks as static quarks $\bar Q$ and computes $\bar Q \bar Q$ potentials in the presence of two lighter quarks $q q$ ($q \in \{ u, d, s \}$) using lattice QCD (see e.g.\ Refs.\ \cite{Detmold:2007wk,Wagner:2010ad,Bali:2010xa,Wagner:2011ev,Brown:2012tm,Bicudo:2015kna}). Then, in a second step, the resulting potentials are inserted into the Schr\"odinger equation to study the dynamics of the heavy $\bar b$ quarks. Using standard techniques from quantum mechanics and scattering theory one can check, whether these potentials are sufficiently attractive to host bound states or resonances, which indicate the existence of $\bar b \bar b q q$ tetraquarks (see e.g.\ Refs.\ \cite{Bicudo:2012qt,Bicudo:2015vta,Bicudo:2016ooe,Bicudo:2017szl}).

At large $\bar Q \bar Q $ separation $r$, the four quarks will form two static-light mesons $\bar Q q$ and $\bar Q q$ and the corresponding potential is equal to the sum of the two meson masses. A $\bar Q \bar Q$ potential in the presence of two lighter quarks $q q$ depends on
\begin{itemize}
\item the light quark flavors (i.e.\ isospin and strangeness),

\item the light quark spins (the static quark spins are irrelevant),

\item parity, which can be related to the types of the mesons (negative parity $B$ and $B^\ast$ ground state mesons and positive parity $B_0^\ast$ and $B_1^\ast$ excitations).
\end{itemize}
Thus, there are quite a number of different channels, which were computed and are discussed in detail in Ref.\ \cite{Bicudo:2015kna}. Some of the corresponding potentials are attractive, others are repulsive, and they differ in their asymptotic values at large $r$.

\begin{figure}[htb]
\begin{center}
\includegraphics[width=0.49\columnwidth]{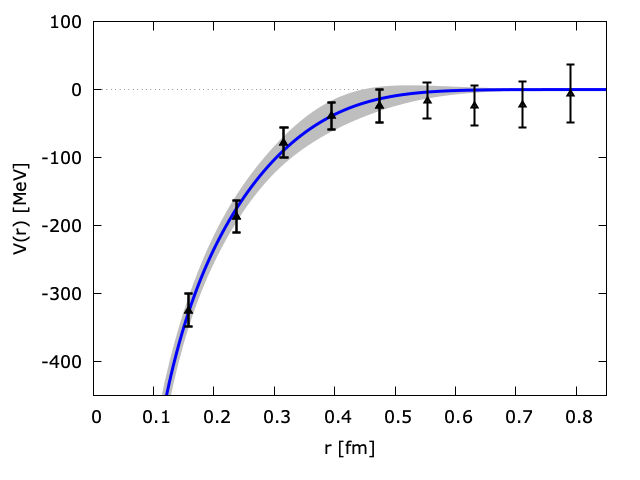}
\includegraphics[width=0.49\columnwidth]{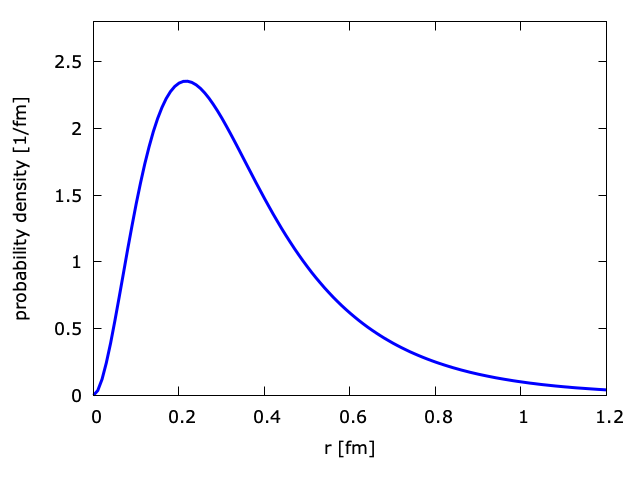}
\end{center}
\caption{\label{FIG001}\textbf{(left)}~Lattice QCD results for the most attractive $\bar Q \bar Q$ potential with quantum numbers $(I,|j_z|,P,P_x) = (0,0,+,-)$ together with the parameterization (\ref{EQN001}). \textbf{(right)}~Radial probability density of the $\bar{b} \bar{b}$ separation $p_r(r) = 4 \pi |R(r)|^2$. (The results shown in the two plots correspond to ensemble B40.24 and are taken from Ref.\ \cite{Bicudo:2012qt}.)}
\end{figure}

To determine $\bar Q \bar Q$ potentials, one has to compute temporal correlation functions of suitably chosen creations operators. One possibility is to use operators of meson-meson type,
\begin{eqnarray}
\label{EQN002} \mathcal{O}_{BB} = 2 (\mathcal{C} \Gamma)_{AB} (\mathcal{C} \tilde{\Gamma})_{CD} \Big(\bar{Q}_C^a(-\mathbf{r}/2) \psi^{(f) a}_A(-\mathbf{r}/2)\Big) \Big(\bar{Q}_D^b(+\mathbf{r}/2) \psi^{(f') b}_B(+\mathbf{r}/2)\Big) ,
\end{eqnarray}
where $\mathcal{C} = \gamma_0 \gamma_2$ is the charge conjugation matrix, $A, B, C, D$ denote spin indices, $a, b$ color indices and $\psi^{(f)}$ represent light quark field operators of flavor $f$. The most attractive potential corresponds to quantum numbers $(I,|j_z|,P,P_x) = (0,0,+,-)$ (for a detailed discussion see Ref.\ \cite{Bicudo:2015kna}) and can be obtained by choosing $\psi^{(f)} \psi^{(f')} = ud - du$, $\Gamma = (1 + \gamma_0) \gamma_5$ and $\tilde{\Gamma} \in \{ (1 + \gamma_0) \gamma_5 \, , \, (1 + \gamma_0) \gamma_j \}$. Lattice data points computed on 2-flavor ETMC gauge link configurations (see Table~\ref{TAB001} and Refs.\ \cite{ETM:2007xow,ETM:2008zte,ETM:2009ztk}) are shown in Figure~\ref{FIG001} (left plot). These results are consistently parameterized by
\begin{eqnarray}
\label{EQN001} V(r) = -\frac{\alpha}{r} \exp\bigg(-\bigg(\frac{r}{d}\bigg)^p\bigg) + V_0
\end{eqnarray}
with $\alpha = 0.293$, $d = 0.356 \, \text{fm}$ and $p = 2.74$ (the constant $V_0$ contains the self energy of the static quarks and is physically irrelevant; within statistical errors $V_0 = 2 m_\text{sl}$, where $m_\textsl{sl}$ is the mass of the lightest static-light meson).

\begin{table}[htb]
\begin{center}
\begin{tabular}{|c|c|c|c|c|c|c|}
\hline
ensemble & $\beta$ & $a$ in $\textrm{fm}$ & $(L/a)^3 \times T/a$ & $\kappa$ & $\mu$ & $m_\textrm{PS}$ in $\textrm{MeV}$ \\
\hline
B40.24 & $3.90$ & $0.079(3)$ & $24^3 \times 48$ & $0.160856$ & $0.004$ & $340(13)$ \\
C30.32 & $4.05$ & $0.063(2)$ & $32^3 \times 64$ & $0.157010$ & $0.003$ & $325(10)$ \\
\hline
\end{tabular}
\caption{\label{TAB001}ETMC gauge link ensembles used in this work (for details see Refs.\ \cite{ETM:2007xow,ETM:2008zte,ETM:2009ztk}).}
\end{center}
\end{table}

When solving the radial Schr\"odinger equation for that potential, 
\begin{eqnarray}
\bigg(\frac{1}{m_b} \bigg(-\frac{d^2}{dr^2} + \frac{L (L+1)}{r^2}\bigg) + V(r) - 2 m_\text{sl}\bigg) R(r) = E R(r)
\end{eqnarray}
($m_b$ denotes the $b$ quark mass, which can be estimated e.g.\ by the mass of the $B$ meson), one finds for orbital angular momentum $L = 0$ a bound state with binding energy $-E = 38(18) \, \text{MeV}$ \cite{Bicudo:2012qt}. The radial probability density of that state, $p_r(r) =  4 \pi |R(r)|^2$, is shown in Figure~\ref{FIG001} (right plot) indicating that the $\bar b \bar b$ separation is typically between $0.1 \, \text{fm}$ and $0.5 \, \text{fm}$. Using the Pauli principle for the $\bar b$ quarks one can conclude that the quantum numbers of the corresponding $\bar b \bar b u d$ tetraquark are $I(J^P) = 0(1^+)$.

% ********************

\section{Structure of the $\bar b \bar b u d$ tetraquark}

To investigate the structure of the $\bar b \bar b u d$ tetraquark, we consider two significantly different creation operators, which both probe the $(I,|j_z|,P,P_x) = (0,0,+,-)$ sector: a meson-meson (or $BB$) operator as given in Eq.\ (\ref{EQN002}) and a diquark-antidiquark (or $Dd$) operator
\begin{eqnarray}
\nonumber & & \hspace{-0.7cm} \mathcal{O}_{Dd} = -\epsilon^{abc} \Big(\psi^{(f) b}_A({\bf 0}) (\mathcal{C} \Gamma)_{AB} \psi^{(f') c}_B({\bf 0})\Big) \\
 & & \hspace{0.675cm} \epsilon^{ade} \Big(\bar{Q}^f_C(-\mathbf{r}/2) U^{fd}(-\mathbf{r}/2;{\bf 0}) (\mathcal{C} \tilde{\Gamma})_{CD} \bar{Q}^g_D(+\mathbf{r}/2) U^{ge}(+\mathbf{r}/2;{\bf 0})\Big)
\end{eqnarray}
with $U$ denoting straight parallel transporters. We choose the same $\psi^{(f)} \psi^{(f')}$, $\Gamma$ and $\tilde{\Gamma}$ as in Eq.\ (\ref{EQN002}). With these two operators we computed the $2 \times 2$ correlation matrix
\begin{eqnarray}
C_{j k}(t) = \Big\langle \mathcal{O}^\dagger_j(t_2) \mathcal{O}_k(t_1) \Big\rangle = \langle \Omega | \mathcal{O}^\dagger_j(t_2) \mathcal{O}_k(t_1) | \Omega \rangle = \langle \Phi_j(t_2) | \Phi_k(t_1) \rangle ,
\end{eqnarray}
where $| \Omega \rangle$ denotes the vacuum and $| \Phi_j \rangle = \mathcal{O}_j | \Omega \rangle$ are meson-meson ($j = BB$) and diquark-antidiquark ($j = Dd$) trial states.

% **********

\subsection{$BB$ and $Dd$ percentages as functions of the $\bar Q \bar Q$ separation for the anti-static-anti-static- light-light system}

\begin{figure}[htb]
\begin{center}
\includegraphics[width=0.32\columnwidth]{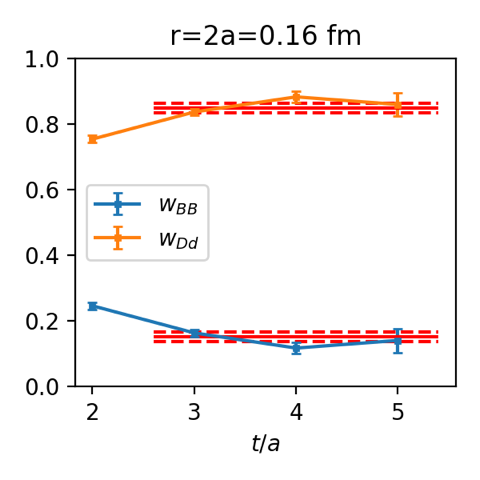}
\includegraphics[width=0.32\columnwidth]{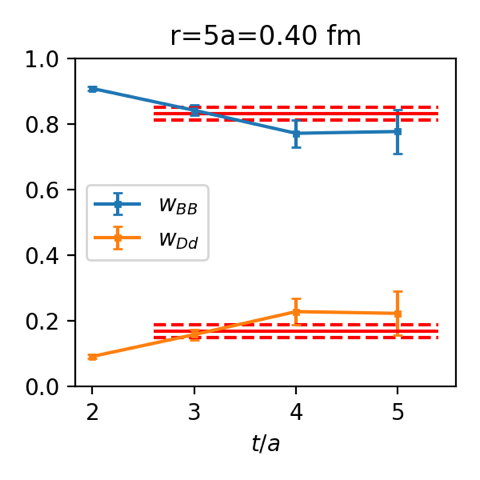}
\includegraphics[width=0.32\columnwidth]{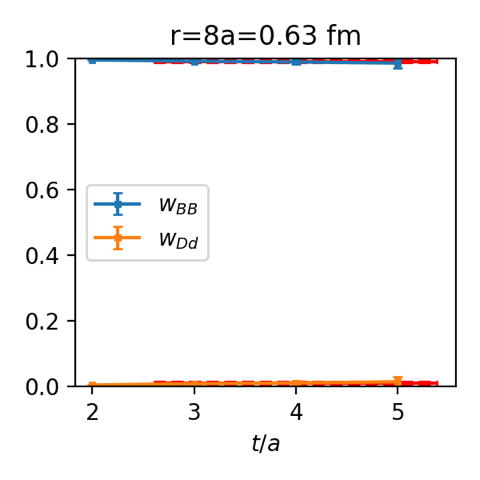}
\end{center}
\caption{\label{FIG002}$w_{BB}$ and $w_{Dd} = 1 - w_{BB}$, the normalized absolute squares of the coefficients of the optimized trial state for several fixed $r$ as functions of $t$ for ensemble B40.24. The horizontal red lines indicate fits of constants $\bar{w}_{BB}$ and $\bar{w}_{Dd}$.}
\end{figure}

In this subsection we focus on the $\bar Q \bar Q u d$ system with static antiquarks with fixed positions. We defined the trial state
\begin{eqnarray}
| \Phi_{b,d} \rangle = b | \Phi_{BB , (1+\gamma_0) \gamma_5} \rangle + d | \Phi_{Dd , (1+\gamma_0) \gamma_5} \rangle
\end{eqnarray}
and determined the coefficients $b$ and $d$ such that the trial state is as similar to the ground state as possible. This amounts to minimizing effective energies
\begin{eqnarray}
V_{b,d}^\text{eff}(r,t) = -\frac{1}{a} \log\bigg(\frac{C_{[b,d] [b,d]}(t)}{C_{[b,d] [b,d]}(t-a)}\bigg) \quad , \quad C_{[b,d] [b,d]}(t) =
\left(\begin{array}{c} b \\ d \end{array}\right)^\dagger_j
C_{j k}(t)
\left(\begin{array}{c} b \\ d \end{array}\right)_k
\end{eqnarray}
with respect to $b$ and $d$. Since the optimization is independent of the normalization and the relative phase of $b$ and $d$, we consider the weights or percentages of $BB$ and $Dd$ defined via
\begin{eqnarray}
w_{BB} = \frac{|b|^2}{|b|^2 + |d|^2} \quad , \quad w_{Dd} = \frac{|d|^2}{|b|^2 + |d|^2} = 1 - w_{BB} .
\end{eqnarray}
For fixed $\bar Q \bar Q$ separation $r$ the percentages $w_{BB}$ and $w_{Dd}$ depend only weakly on $t$ as shown in Figure~\ref{FIG002} for selected separations. To fully eliminate the $t$ dependence, we fit constants $\bar{w}_{BB}(r)$ and $\bar{w}_{Dd}(r)$ to the lattice data points $w_{BB}(r,t)$ and $w_{Dd}(r,t)$ for fixed $r$, but several $t$.

\begin{figure}[htb]
\begin{center}
\includegraphics[width=0.49\columnwidth]{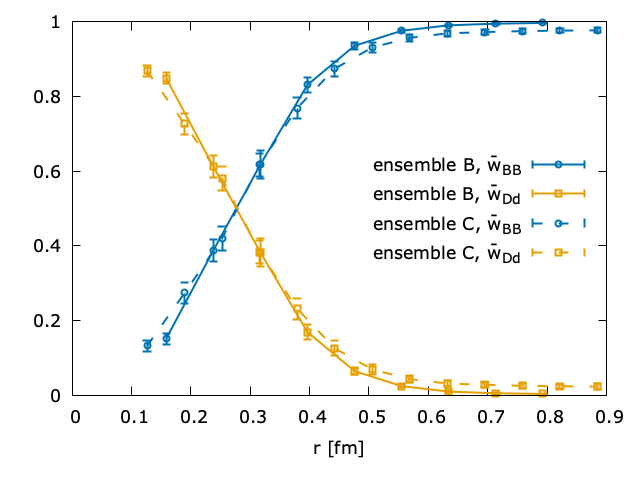}
\end{center}
\caption{\label{FIG003}$\bar{w}_{BB}$ and $\bar{w}_{Dd} = 1 - \bar{w}_{BB}$, the normalized absolute squares of the coefficients of the optimized trial state, as functions of $r$ for both ensembles.}
\end{figure}

In Figure~\ref{FIG003} we show the percentages $\bar{w}_{BB}$ and $\bar{w}_{Dd}$ as functions of the $\bar{Q} \bar{Q}$ separation $r$ for the two ensembles B40.24 and C30.32. For $r \ltapprox 0.2 \, \text{fm}$ there is clear diquark-antidiquark dominance. For $0.2 \, \text{fm} \ltapprox r \ltapprox 0.3 \, \text{fm}$ diquark-antidiquark dominance turns into meson-meson dominance. For $0.5 \, \text{fm} \ltapprox r$ the system is essentially a pair of static-light mesons. There is no significant difference between the two ensembles and our results for $\bar{w}_{BB}$ and $\bar{w}_{Dd}$ seem to be essentially independent of the lattice spacing $a$.

As an alternative to $\bar{w}_{BB}$ and $\bar{w}_{Dd}$ one can also study eigenvector components obtained from a standard generalized eigenvalue problem. Results on the $BB$ and $Dd$ percentages are very similar. For details see Ref.\ \cite{Bicudo:2021qxj}.

% **********

\subsection{$BB$ and $Dd$ percentages for the $\bar b \bar b u d$ tetraquark}

The total meson-meson and diquark-antidiquark percentages of the $\bar b \bar b u d$ tetraquark can be obtained by numerically solving the integrals
\begin{eqnarray}
\%BB = \int dr \, p_r(r) \bar{w}_{BB}(r) \quad , \quad \%Dd = \int dr \, p_r(r) \bar{w}_{Dd}(r) = 1 - \%BB ,
\end{eqnarray}
where $p_r(r) = 4 \pi |R(r)|^2$ is the radial probability density discussed in section~\ref{SEC001} and shown in Figure~\ref{FIG001} (right plot). We find $\%BB = 0.58$ and $\%Dd = 0.42$. These results indicate that the $\bar b \bar b u d$ tetraquark with quantum numbers $I(J^P) = 0(1^+)$ is a linear superposition of a meson-meson system and a diquark-antidiquark system with slight meson-meson dominance. This is supported by a recent full lattice QCD study of the same system using four quarks of finite mass \cite{Leskovec:2019ioa,Pflaumer:2021ong}.

% ********************

\section*{Acknowledgements}

We acknowledge useful discussions with A.\ Ali, C.\ Hanhart, J.\ K\"amper, M.\ Pflaumer and G.\ Schierholz.

M.W.\ acknowledges support by the Heisenberg Programme of the Deutsche Forschungsgemeinschaft (DFG, German Research Foundation) -- project number 399217702. P.B.\ thanks the support of CeFEMA under the contract for R\&D Units, strategic project No.\ UID/CTM/04540/2019, and the FCT project Grant No.\ CERN/FIS-COM/0029/2017.

Calculations on the GOETHE-HLR and on the on the FUCHS-CSC high-performance computers of the Frankfurt University were conducted for this research. We would like to thank HPC-Hessen, funded by the State Ministry of Higher Education, Research and the Arts, for programming advice.

% ********************

% ********************

\end{document}